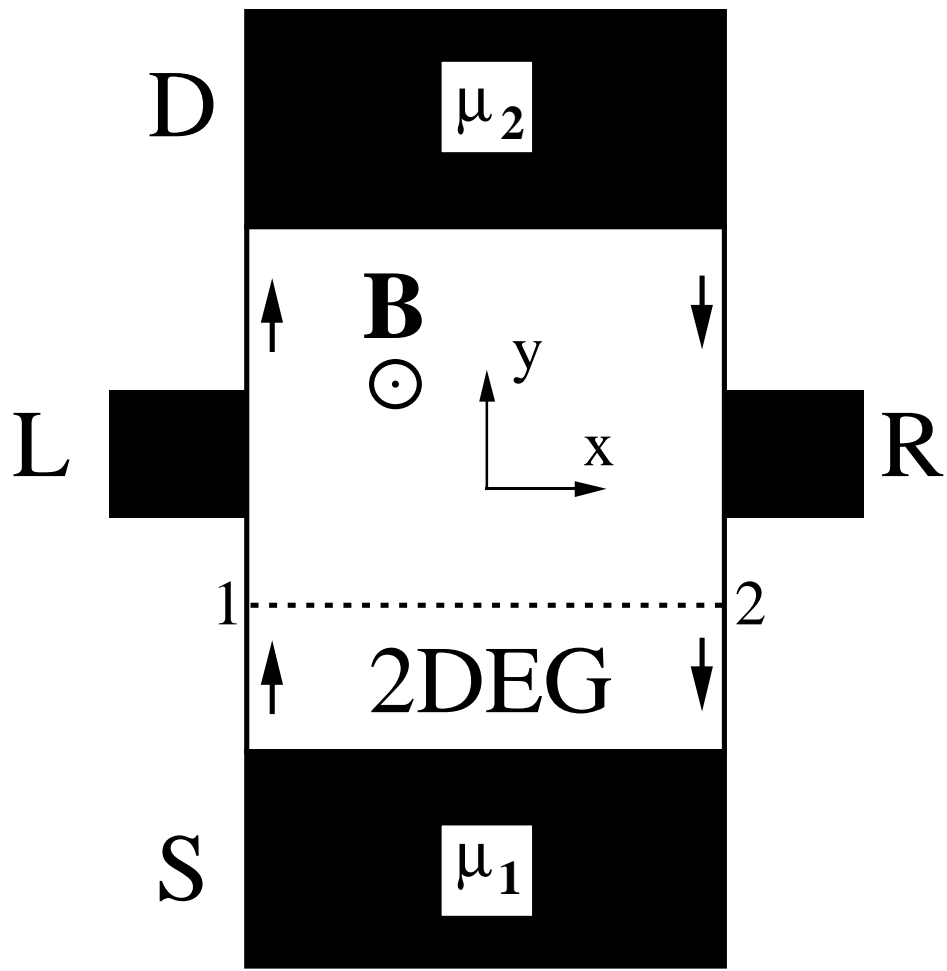

Figure 1

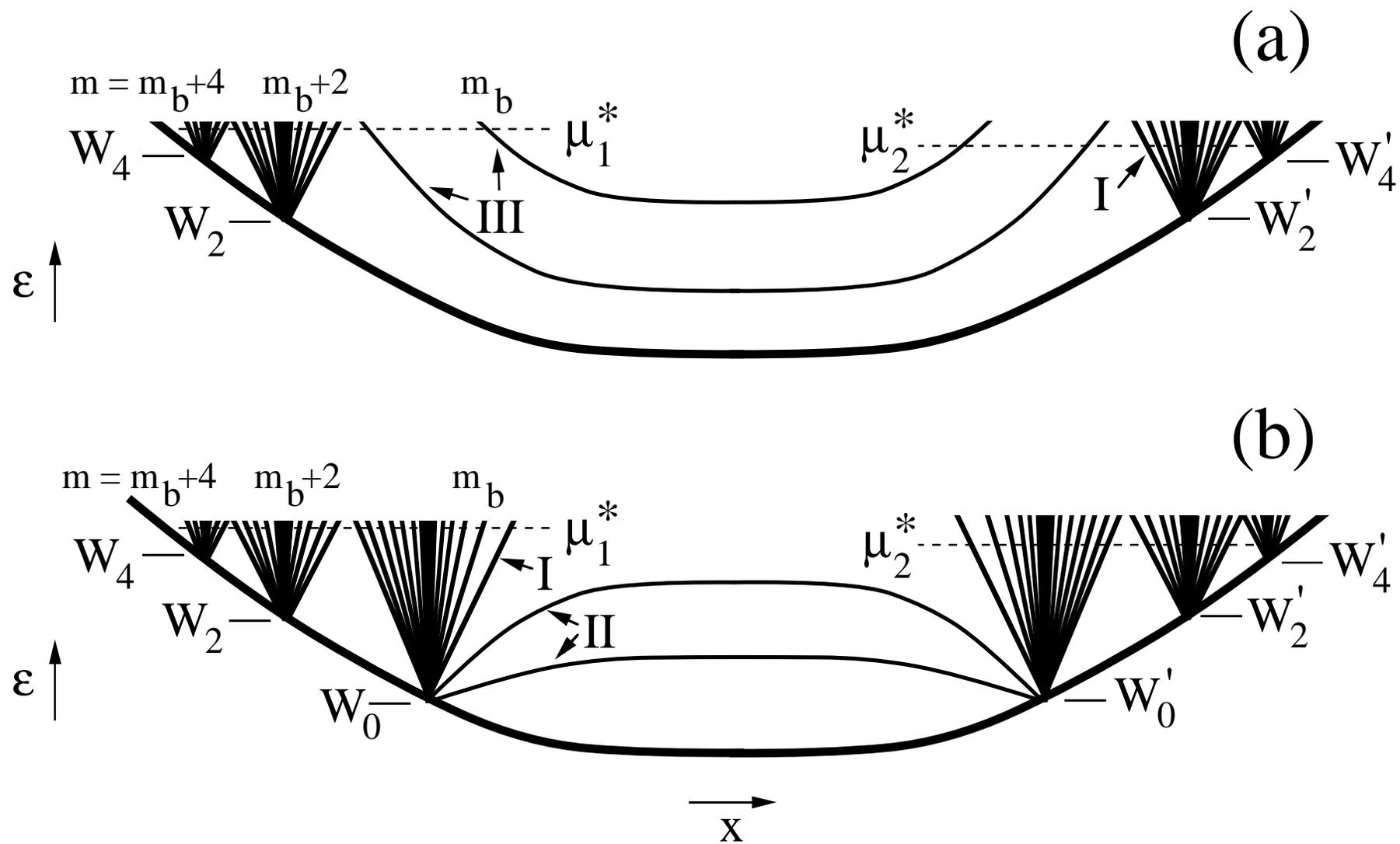

Figure 2

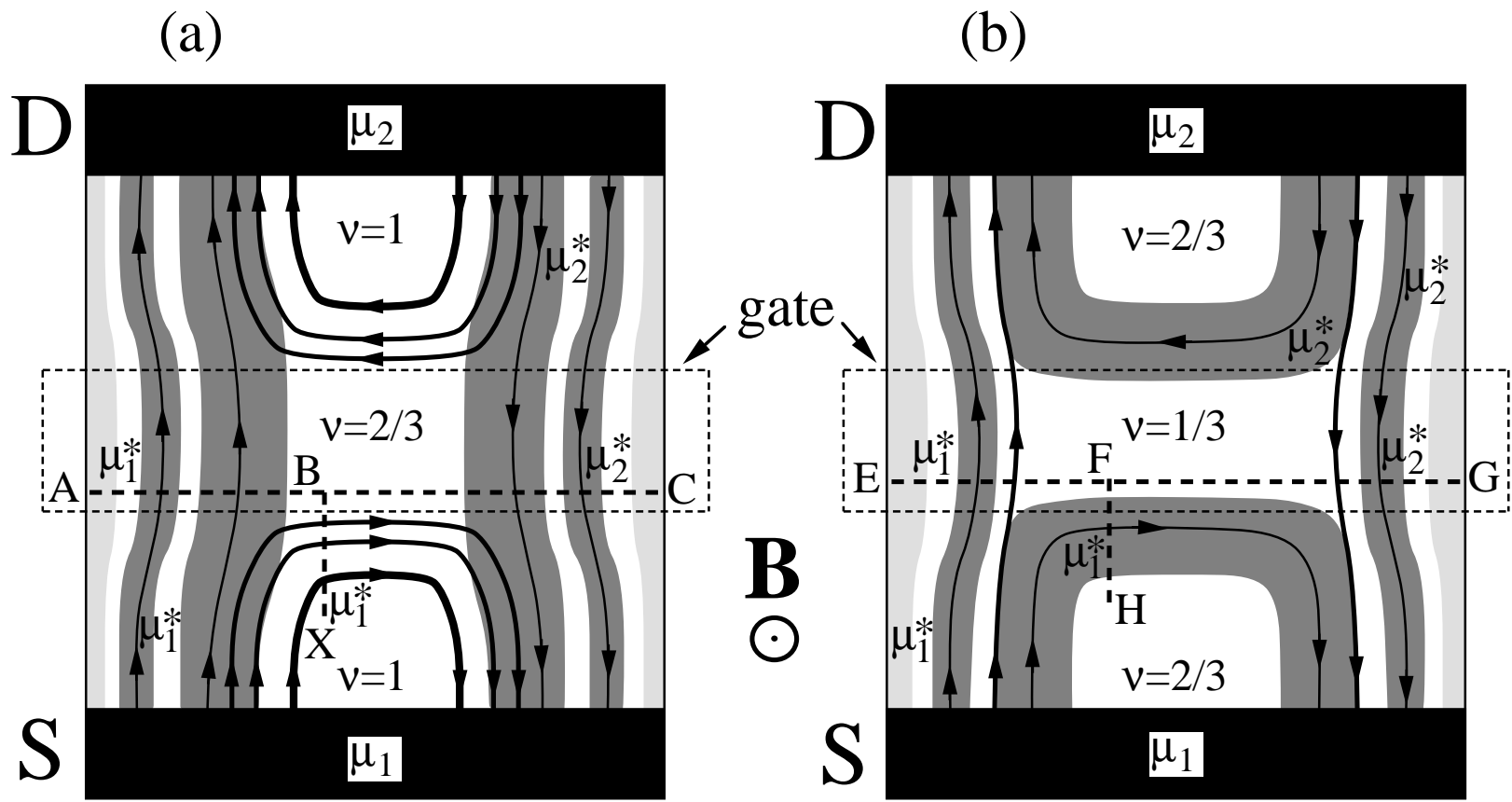

Figure 3

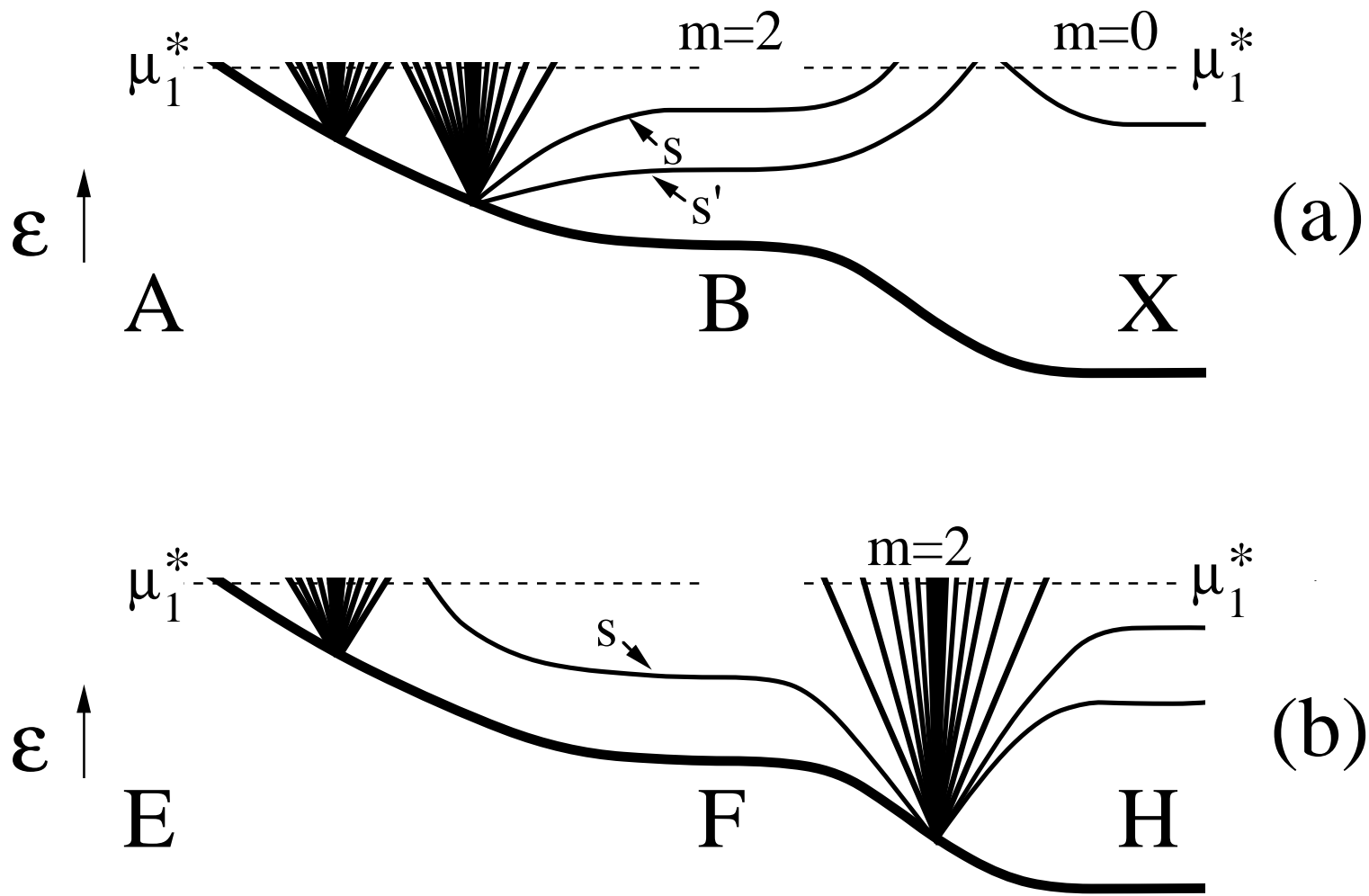

Figure 4

Figure 5

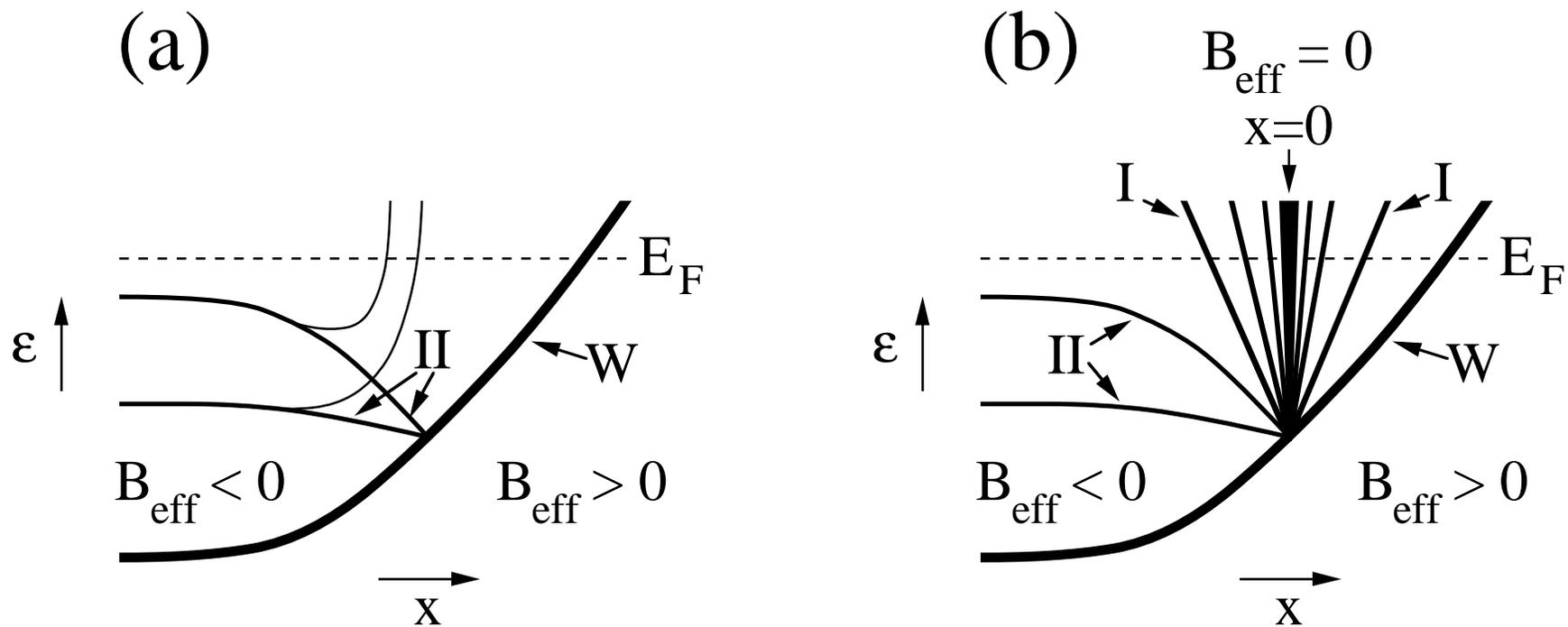

Figure 6

# Composite Fermions, Edge Currents and the Fractional Quantum Hall Effect

George Kirczenow and Brad L. Johnson

*Department of Physics, Simon Fraser University,*
*Burnaby, British Columbia,*
*Canada, V5A 1S6*

We present a mean field theory of composite fermion edge states and their transport properties in the fractional and integer quantum Hall regimes. Slowly varying edge potentials are assumed. It is shown that the effective electro-chemical potentials of composite fermions at the edges of a Hall bar differ, in general, from those of electrons, and an expression for the difference is obtained. Composite fermion edge states of three different types are identified. Two of these types have no analog in previous theories of the integer or fractional quantum Hall effect. The third type includes the usual integer edge states. The direction of propagation of the edge states is consistent with experimental observations. The present theory yields the experimentally observed quantized Hall conductances at the bulk Landau level filling fractions $\nu = p/(mp \pm 1)$, where m=0, 2, 4, and p = 1, 2, 3, ... It also explains the results of experiments that involve conduction across smooth potential barriers and through adiabatic constrictions, and of experiments that involve selective population and detection of edge channels in the fractional quantum Hall regime. The relationship between the present work and Hartree theories of composite fermion edge structure is discussed.

## 1. Introduction

The discovery that the Hall conductance of two-dimensional electron systems is quantized at low temperatures in integer[1] and fractional[2] multiples of $e^2/h$ has stimulated the emergence of many remarkable theoretical ideas.

The integer quantum Hall effect occurs when the Fermi level lies in a mobility gap between the bulk Landau levels of a two-dimensional electron system. It was first explained by Laughlin using a gauge invariance argument.[3] Subsequently, Streda, Kucera and MacDonald,[4] Jain and Kivelson[5] and Büttiker[6] proposed an alternate point of view in which it is explained on the basis of the Büttiker-Landauer theory of one-dimensional conduction,[7] within the framework of magnetic edge states introduced by Halperin.[8]

Unlike the integer quantum Hall effect, the fractional quantum Hall effect is believed to be inherently a many-body phenomenon. As was shown by Laughlin,[9] it is due to the formation of incompressible states of a partly filled Landau level at the filling fractions 1/3, 1/5, 1/7 ... A hierarchical generalization of Laughlin's states was able to explain the occurrence of the fractional quantum Hall effect at other fractional fillings.[10] A different generalization of Laughlin's states was introduced by Jain.[11] Jain's picture of the fractional quantum Hall effect can be thought of in terms of a singular gauge transformation of the many-electron Hamiltonian and eigenstates, that is related to the transformations from fermion to anyon and boson representations that had been introduced earlier.[12] This gauge transformation[11,13,14] results in a tube of fictitious magnetic flux (carrying an even number of flux quanta) being attached to each electron. In the new gauge, the electrons, together with the flux tubes attached to them, obey Fermi statistics and are thus known as "composite fermions." The simplest description of composite fermion systems is a mean field theory described by Jain,[11] Lopez and Fradkin[13] and Halperin, Lee and Read.[14] In the mean field theory, the composite fermions interact with an effective magnetic field that is the sum of the true magnetic field and a fictitious magnetic field that is the spatial average of the gauge flux carried by the composite fermions. The fractional quantum Hall effect is observed at magnetic fields and electron densities at which an integer number of Landau levels is occupied by the composite fermions in this effective magnetic field. Thus the fractional quantum Hall effect can be viewed as a manifestation of the integer quantum Hall effect for composite fermions.[11] The integer quantum Hall effect also fits into this picture, as a special case.[11] In addition to unifying the integer and fractional quantum Hall effects, the composite fermion theory explains[11] the relative stabilities of the different fractional quantum Hall phases that are observed experimentally.[2,15] It also explains the scaling with magnetic field of the energy gaps that are observed in the fractional quantum Hall regime.[16,17,18] Another consequence of the composite fermion theory, demonstrated theoretically by Halperin, Lee and Read,[14] is that at the Landau level filling factor $\nu = 1/2$, where the effective magnetic field vanishes, the lowest spin-polarized electron Landau level has many features in common with a system of electrons at zero magnetic field. For $\nu \approx 1/2$ it behaves somewhat like a system of quasi-classical electrons at very low magnetic fields. These remarkable predictions have been confirmed experimentally.[19,20,21]

Working in the Corbino geometry, Jain demonstrated that composite fermion theory yields the experimentally observed fractionally quantized Hall conductances. His proof[11] was a generalization of the arguments that Laughlin[3] had used earlier to explain the integer quantum Hall effect. However, because of the very close relationship between the integer and fractional quantum Hall effects in the composite fermion picture, one might hope to also construct a composite fermion theory of transport in the fractional quantum Hall regime that is somewhat closer in spirit to the theories of edge channel transport [4–8] in the integer quantum Hall regime. This is done, at the level of mean field theory, in the present paper.

In Section 2, we review briefly the main ideas underlying the mean field theory of unbounded two-dimensional composite fermion systems, and then examine some of their implications for transport classically. In Section 3, we demonstrate that the effective electro-chemical potentials of the composite fermions at the edges of a finite two-dimensional system that carries a current, can differ from the electro-chemical potentials of the corresponding electrons at the same edges. We also show how these quantities are related, using the mean field ideas outlined in Section 2, and the fact that the local charge density is gauge-invariant.

In Section 4, we consider the nature of the composite fermion states near the edges of a Hall bar and their transport properties, within mean field theory. We assume that the electrostatic potential is slowly varying spatially, and that quasi-equilibrium exists at each edge. In the present theory, the composite fermion edge states are of three distinct types: Type I edge states, although they cross the composite fermion Fermi level, are "silent" modes. We show that the electric current that they carry does not change when the edge electro-chemical potential changes, provided that quasi-equilibrium at that edge is maintained. Type II edge states lie below the composite fermion Fermi level, and are separated from it by a finite gap when the system exhibits a fractional quantum Hall plateau. However these modes *do* respond to a change in the edge electro-chemical potential with a change in the current they carry. Thus the properties of the Type I and Type II edge states are fundamentally unlike those of the ordinary magnetic edge states that appear in the standard theories of the integer quantum Hall effect. Type III edge states cross the composite fermion Fermi level and do respond to a change in the edge electro-chemical potential with a change in the current that they carry. The usual edge states of the integer quantum Hall effect[8] are a special case of Type III edge states. The direction of propagation of the edge states described in this article is in agreement with experiment. We also show how the correct fractionally quantized Hall conductance of a Hall bar connected to ideal current and voltage contacts follows from the properties of these edge states and from the relationship between the electron and composite fermion edge electro-chemical potentials that is established in Section 3.

In Section 5, we discuss the application of the theory of composite fermion edge channels developed in Section 4 to conductors with smooth potential barriers and adiabatic constrictions, and find that the results agree with the relevant experiments. The theory is shown to explain the results of experiments that involve selective population and detection of fractional edge channels in Section 6. Our conclusions are summarized in Section 7. In the Appendix we discuss the relationship between the present work and Hartree theories of composite fermion edge structure, the direction of propagation of the composite fermion edge states, and the possible role of the many-body effects that result in the divergence of the composite fermion effective mass that occurs when the effective magnetic field vanishes.

## 2. Mean Field Theory of Unbounded Composite Fermion Systems

In mean field theory, the interactions between composite fermions that are due to the vector potentials associated with the tubes of gauge flux attached to the electrons, are replaced by interactions with a fictitious average magnetic field and a fictitious electric field:

As discussed by Jain,[11] Lopez and Fradkin[13] and Halperin, Lee and Read,[14] the fictitious magnetic field

$$\mathbf{B}_g = -mn\hat{\mathbf{B}}h/e = -m\nu\mathbf{B} \tag{1}$$

is equal to the average gauge flux per unit area, and points in the direction opposite to that of the real magnetic field $\mathbf{B}$. Here $m$ is the (even) number of quanta of gauge flux carried by each composite fermion, $n$ is the two-dimensional electron density, which is assumed to be uniform, $\nu = nh/(eB)$ is the Landau level filling parameter, and $\hat{\mathbf{B}}$ is the unit vector in the direction of the true magnetic field $\mathbf{B}$, which is normal to the plane containing the electrons. Thus the composite fermions are taken to interact with an effective magnetic field

$$\mathbf{B}_{\text{eff}} = \mathbf{B} + \mathbf{B}_g \tag{2}$$

It follows[11,13,14] that at filling fractions $\nu = p/(mp \pm 1)$ of the lowest spin-polarized electron Landau level, for positive integers $p$, exactly $p$ Landau levels are filled by composite fermions in the *effective* magnetic field $\mathbf{B}_{\text{eff}}$. The fractional quantum Hall effect that is observed at the electron filling fractions $\nu = p/(mp \pm 1)$ is therefore associated[11] with integer filling of the composite fermion Landau levels.

If a composite fermion system carries an electric current, the tubes of gauge flux that are attached to the electrons drift through the system, and this drifting gauge flux implies[22,23,24] that a fictitious electric field

$$\mathbf{E}_g = -(\mathbf{J} \times \hat{\mathbf{B}})\, mh/e^2, \tag{3}$$

the analog of the electric field in a moving solenoid, is also experienced by the composite fermions. Here **J** is the two-dimensional electric current density. Similar gauge electric fields have also been discussed in the context of boson[25] and anyon[26,27] Chern-Simons theories.

The importance of including the gauge electric field $\mathbf{E}_g$ in any consistent composite fermion mean field theory of transport becomes clear if one considers the case of a half-filled Landau level, $\nu=1/2$. In this case, for the standard choice of two quanta of gauge flux per composite fermion ($m=2$), equations (1) and (2) yield $B_{eff}=0$. But at zero magnetic field, the Hall resistance of an electron system is zero, whereas experiment shows that at $\nu=1/2$ the Hall resistance is $2h/e^2$. Thus simply treating composite fermions as if they were electrons moving in an effective magnetic field $B_{eff}$ is incorrect. However, according to equation (3), the system of composite fermions moving in a zero *effective* magnetic field generates the transverse *electric* field $\mathbf{E}_g$ by virtue of its motion, and thus exhibits a Hall voltage proportional to the net electric current, and the correct Hall resistance $2h/e^2$ is thus obtained.

More generally, combining equations (1) and (3), one finds that $\mathbf{E}_g = -\langle\mathbf{v}\rangle\times\mathbf{B}_g$ where $\langle\mathbf{v}\rangle = -\mathbf{J}/(ne)$ is the electron (or composite fermion) average drift velocity. Thus, in classical language, the fictitious electric field $\mathbf{E}_g$ exerts a force $-e\mathbf{E}_g$ equal to $e\langle\mathbf{v}\rangle\times\mathbf{B}_g$ on the composite fermions which exactly cancels the *average* magnetic force $-e\langle\mathbf{v}\rangle\times\mathbf{B}_g$ due to the fictitious magnetic field $\mathbf{B}_g$. Thus $-e\mathbf{E}_g - e\langle\mathbf{v}\rangle\times\mathbf{B}_{eff} = -e\langle\mathbf{v}\rangle\times\mathbf{B}$. That is, the *average* Lorentz force experienced by the composite fermions is the same as that experienced by ordinary electrons. We note, however, that since in most experiments the electric current is weak and $|\langle\mathbf{v}\rangle| \ll v_F$, the *individual* composite fermion semiclassical trajectories are still determined mainly by the effective magnetic force $-e\mathbf{v}_F\times\mathbf{B}_{eff}$, and therefore the classical trapping and focussing arguments discussed in Refs. 20 and 21 remain valid.

While the gauge electric field $\mathbf{E}_g$ compensates the average classical Lorenz force $-e\langle\mathbf{v}\rangle\times\mathbf{B}_g$ associated with the fictitious magnetic field $\mathbf{B}_g$, the fictitious scalar potential associated with $\mathbf{E}_g$ in composite fermion theory is not cancelled in this way, since a magnetic field is not associated with a scalar potential. It will be shown in the following Section, that the presence of this scalar potential associated with $\mathbf{E}_g$ implies that the effective *electro-chemical* potentials of composite fermions at the edges of a Hall bar differ, in general, from those of the corresponding electrons. This difference is very important and must be taken into account when comparing the predictions of mean field theory of composite fermion quantum edge channels with the results of transport measurements on Hall bars connected to current and voltage leads as will be seen in Section 4.

## 3. Composite Fermion Electro-Chemical Potentials

Consider now a wide but finite two-dimensional conductor, connected to ideal electron reservoirs with well-defined electrochemical potentials as depicted in Fig.1. The magnetic field **B** is perpendicular to the electron plane. S and D are the source and drain contacts for electrons and are maintained at electrochemical potentials $\mu_1$ and $\mu_2$, respectively. The directions of the electron flow at the edges of the sample are shown by arrows. L and R are Hall voltage contacts that draw no net current. Thus on a quantum Hall plateau, the electrochemical potentials of L and R are equal to $\mu_1$ and $\mu_2$, respectively. In fact, on a quantum Hall plateau, the electro-chemical potentials of the electrons are equal to $\mu_1$ and $\mu_2$ everywhere along the left and right hand edges, respectively, of the Hall bar shown in Fig.1.

It is important to realize, however, that the effective electro-chemical potential of the composite fermions may be different from that of the electrons at the same edge. This can be seen as follows: The transformation from electrons to composite fermions is a gauge transformation, and it therefore preserves the square amplitude of the many-particle wave function. This implies that the local density *n* of composite fermions is equal everywhere to the local density of electrons. The local electric field experienced by the composite fermions can therefore be viewed as consisting of two parts: **The first part** is the electrostatic field that can be obtained by solving Poisson's equation (given the local charge density everywhere in the system). It is *the same* as the electric field that is experienced by the electrons, because the local electron density is equal everywhere to the local composite fermion density. **The second part** is the fictitious gauge electric field given by equation (3). It exists *only* in the representation of composite fermions, and not in that of electrons. Thus, if it is not zero, it results in the composite fermions having a different effective electrostatic potential energy and hence a different electro-chemical potential from that of the electrons. Let $\mu^*$ be the effective *composite fermion* electro-chemical potential that corresponds to the *electron* electro-chemical potential $\mu$. We are interested in the differences between the effective composite fermion electro-chemical potentials at different edges *i* and *j* of the Hall bar. These differences can be expressed in terms of the *electron* electro-chemical potentials at the respective edges and the current flowing between the edges as

$$\mu_i^* - \mu_j^* = \mu_i - \mu_j + \int_j^i e\mathbf{E}_g . d\mathbf{r} = \mu_i - \mu_j - \int_j^i (mh/e) \mathbf{J}.\hat{\mathbf{B}} \times d\mathbf{r}. \qquad (4)$$

$\int_j^i \mathbf{J}.\hat{\mathbf{B}} \times d\mathbf{r}$ is just the net electric current flowing across the path connecting the two edges. Notice that the gauge electric field $\mathbf{E}_g$ that enters equation (4) is *unscreened* because the effects of the electrostatic fields arising from the *exact* charge distribution (which is the same for both electrons and composite fermions) are already taken fully into account in the difference $\mu_i - \mu_j$ of the *electron* electro-chemical potentials that appears on the RHS.

It is also important to note that equation (4), is an expression that relates *thermodynamic* quantities and assumes local quasi-equilibrium. It therefore only applies to electro-chemical potential differences between *different* edges of a sample, and should not to be taken to imply variations of electro-chemical potentials over short length scales. The relationship between the electro-chemical potentials of the different composite Fermion edge channels at any *single* edge is determined not by equation (4) but by the processes by which carriers are injected into the 2D system. In this article we assume that the contacts are ideal in the sense that there is quasi-equilibrium between the *composite Fermion* edge channels originating from any particular contact. However, in some experiments, channels originating at *different* contacts converge to the *same* edge of the sample. Such non-equilibrium situations are discussed in detail in Section 5 and Section 6.

To summarize: The *electron* electro-chemical potential difference between the edges of a Hall bar can be viewed as being due to the piling up of charge along the edges of the current-carrying conductor in a magnetic field. This charge pile up is the same for composite fermions as for electrons, but the composite fermions experience an *additional* gauge electric field $\mathbf{E}_g$, and an associated additional potential difference across the Hall bar. This means that the effective electro-chemical potential difference across the Hall bar *for composite fermions* differs from that for electrons by a term $\int_j^i e\mathbf{E}_g . d\mathbf{r}$. The implications of this difference for the mean field theory of composite fermion edge transport are made clear in the following Section.

# 4. Edge States and Quantization of the Hall Conductance in Composite Fermion Mean Field Theory

Let us suppose that the electrostatic potential experienced by the electrons is *slowly varying* on the scale of the magnetic length, as the edge of the Hall bar is approached. As the edge is approached, the local electron density $n$ and Landau level filling parameter $\nu$ decrease, and $B_{\text{eff}}$ also varies because it depends on the local density according to (1) and (2). $B_{\text{eff}}$ vanishes when $\nu=1/m$, for even integers $m$. In this paper, we assume that for $\nu$ in the vicinity of $1/m$, the mean field composite fermion Landau level energies are given by

$$\varepsilon_{m,r} \approx \left(r + \frac{1}{2}\right)\hbar e |B_{\text{eff}}|/m^* + W \qquad (5)$$

Here $r = 0, 1, 2, ...$ and $W$ is the composite fermion effective potential energy which depends on position. Note that $W$ includes a contribution associated with the gauge electric field. $m^*$ is a phenomenological composite fermion effective mass. It has been predicted theoretically[14] that the effective mass should diverge when $B_{\text{eff}} = 0$. The experimental data appears to support this, but suggests that the divergence may be much stronger than predicted.[16,17,18,28]

Equation (5) is generally believed to provide a good description of the composite fermion energy level structure as a function of magnetic field in *uniform* systems. It is widely used to interpret experimental data. But some recent theories[29,30] have suggested that the behavior of $\varepsilon_{m,r}$ near the edges of composite fermion systems may differ qualitatively from that predicted by equation (5). However, those theories do not take into account many-body effects that give rise to the divergence of the effective mass at $B_{\text{eff}} = 0$, which we believe to be important, and they yield a direction of propagation for the composite fermion edge states in the $\nu=2/3$ regime that is the opposite to what is observed experimentally. We discuss the relationship between those theories and the present work and also the significance of the divergence of the effective mass in the Appendix.

In Fig.2, the thinner curves show $\varepsilon_{m,r}$ (equation (5)) schematically as a function of position $x$ along the dashed line in Fig.1; the thicker curves represent $W$. Fig.2(a) and Fig.2(b) show cases where, deep in the interior of the 2DEG, $\mathbf{B}_{\text{eff}}$ is parallel and anti-parallel to the real magnetic field $\mathbf{B}$, respectively. Each "fan" of lines representing $\varepsilon_{m,r}$ corresponds to a particular value of $m$ as indicated, beginning with $m_b$ in the bulk. Other fans of levels that may be obtained from these by particle-hole transformations[11] and are centered at filling fractions $\nu = 1-1/m$ can also occur, but are not shown. They are not considered further in the present paper, where we will be concerned only with the predominant fractional quantum Hall phases that correspond to $\nu = p/(mp \pm 1)$, and the composite fermion Landau levels that are associated with these fractions.

In order to study the transport characteristics of the conducting channels associated with these composite fermion mean field Landau levels, consider the vicinity of the dashed line in Fig.1 where the system is translationally invariant in the $y$-direction. Let us introduce a mean field effective vector potential

$$\mathbf{A}_{\text{eff}} = (0, \int^x B_{\text{eff}}(u)\, du, 0) \qquad (6)$$

for which $\nabla \times \mathbf{A}_{\text{eff}} = \mathbf{B}_{\text{eff}}$. The mean field single-particle composite fermion eigenstates are then solutions of the effective Schrödinger equation

$$\left(\frac{1}{2m^*}\left(\frac{\hbar}{i}\nabla + e\mathbf{A}_{\text{eff}}\right)^2 + W(x)\right)\Psi_{k,m,r} = \varepsilon_{m,r}(k)\Psi_{k,m,r} \qquad (7)$$

and are of the form
$$\Psi_{k,m,r} = e^{iky} X_{k,m,r}(x). \tag{8}$$

As in the edge channel theories of the integer quantum Hall effect,[4–8] the electric current carried by a particular channel $(m,r)$ is then

$$I_{m,r} = -\frac{e}{h}\int \frac{\partial \varepsilon_{m,r}}{\partial k} f_{m,r}(k)\,dk = -\frac{e}{h}\int d\varepsilon_{m,r} = -\frac{e}{h}\Delta\varepsilon_{m,r} \tag{9}$$

where $f_{m,r}(k)$ is the number of composite fermions occupying the state $\Psi_{k,m,r}$. Note that the charge of a composite fermion is $-e$, the same as the charge an electron, because gauge transformations preserve charge. The second equality is obtained by assuming that $f_{m,r}(k) = 1$ or $0$ depending on whether the state $\Psi_{k,m,r}$ is occupied or not. We note that $\partial\varepsilon_{m,r}/\partial x$ is positive for some of the modes $(m,r)$ depicted in Fig.2 and negative for others. However, analysis of equations (6) and (7) (see the Appendix) shows that $\partial\varepsilon_{m,r}/\partial k$ is positive for all of the modes shown in the left half of Fig.2 and negative for all of the modes shown in the right half. Thus all of the composite fermion modes that are shown in Fig.2 travel in the directions indicated by the arrows in Fig.1, the same directions as for non-interacting electrons.[31]

Let us suppose that the composite fermion states $\Psi_{k,m,r}$ are fully populated by the electron reservoirs up to a composite fermion electro-chemical potential $\mu_1^*$ for modes flowing from reservoirs S and L, and up to $\mu_2^*$ for modes flowing from reservoirs D and R in Fig.1. Under these quasi-equilibrium conditions, we now calculate the change $dI_{m,r}$ in the current carried by mode $m,r$ across the dashed line in Fig.1 (and also the change in the total current $dI$), when the electro-chemical potential $\mu_1^*$ is changed by a small amount $d\mu_1^*$, and $\mu_2^*$ is changed by $d\mu_2^*$.

There are composite fermion edge modes of three different types shown in Fig.2. They and their responses to changes in the edge electro-chemical potentials will now be discussed:

**Type I edge modes** are those that correspond to mean field composite fermion Landau levels that are confined entirely to the vicinity of a particular edge. In Fig.2 these are the Landau levels that begin at the apex of a "fan" (where $B_{\text{eff}} = 0$) and pass through the Fermi level at an energy equal to the composite fermion electro-chemical potential $\mu_i^*$ for that edge. According to equation (9), the current carried by an edge mode of this type belonging to edge 1 (see Fig.1) changes by

$$dI_{m,r} = -\frac{e}{h}\left(d\mu_1^* - dW_\alpha\right) \tag{10}$$

in response to a change $d\mu_1^*$ of the composite fermion electro-chemical potential. Here $W_\alpha$ is the composite fermion potential energy where $B_{\text{eff}} = 0$ for the given mode, as indicated in Fig.2. In a quasi-equilibrium system with a well-defined composite fermion edge electro-chemical potential $\mu_1^*$, and if $W$ is slowly varying with position, it is reasonable to suppose that, at any given real magnetic field $B$, $\mu_1^*-W$ (the local Fermi energy) depends *only on the local composite fermion* (or electron) *density n*. But now recall that $W_\alpha$ in equation (10) is evaluated where $B_{\text{eff}} = 0$, and that $B_{\text{eff}} = 0$ occurs at a particular value $eB/(hm)$ of the density $n$ where $\nu = 1/m$.[38] This value of the density is independent of the electro-chemical potential $\mu_1^*$. Therefore $\mu_1^*-W_\alpha$ is independent of $\mu_1^*$. That is, $W_\alpha$ exactly tracks the effective composite fermion edge electro-chemical potential. We thus obtain that

$$dW_\alpha = d\mu_1^*. \tag{11}$$

It then follows from equation (10) that the change $dI_{m,r}$ in the current carried by a Type I composite fermion edge mode, that occurs in response to a change $d\mu_1^*$ of the edge electro-chemical potential, is zero. A similar result holds for Type I edge modes at edge 2 in Fig.1. That is, Type I edge modes *do not contribute to the current response* (or to the Hall conductance) of a Hall bar with well defined quasi-equilibrium composite fermion electro-chemical potentials at its edges. In this sense they are "silent" modes, entirely unlike the edge modes that figure in theories of the integer quantum Hall effect.

We emphasize, however, that Type I edge modes are quite real and can, under suitable conditions, transport charge. In cases where edge state quasi-equilibrium is *not* achieved, they can exhibit a non-zero electric current response to a change of reservoir chemical potential. Thus they can be detected experimentally, as will be discussed in Section 5 and Section 6. They also play an important role in electron transport in 2D conductors with smooth potential barriers or adiabatic constrictions. This will be explained in Section 5.

**Type II edge modes** are those that correspond to mean field composite fermion Landau levels that begin at the apex of a fan where $B_{\mathrm{eff}} = 0$, and continue into the interior of the two-dimensional conductor, always staying below the Fermi level. Examples are the two inner-most modes of Fig.2(b). We now consider the current response of these modes to changes in the composite fermion electro-chemical potential $\mu_1^*$, again assuming quasi-equilibrium at the edges, and that $W$ is slowly varying.

In this case equation (9) takes the form

$$dI_{m,r} = -\frac{e}{h}(dW'_0 - dW_0) = \frac{e}{h}d\left(\mu_1^* - \mu_2^*\right) \tag{12}$$

where the second equality is obtained using equation (11), which here takes the form $dW_0 = d\mu_1^*$, and its analog $dW'_0 = d\mu_2^*$ which holds at the right hand edge in Fig.2(b). If there are $p$ Type II modes present ($p$ is also the number of occupied composite fermion Landau levels in the interior of the Hall bar) then the change in the total current is

$$dI = \sum_r dI_{m,r} = p\frac{e}{h}d\left(\mu_1^* - \mu_2^*\right), \tag{13}$$

since the Type I modes are silent, and do not contribute to the current response, as has been explained above. Notice that, in this case, the current response $dI$ comes entirely from modes that are separated by a gap from the Fermi level, another unique feature of composite fermion theory, not found in the theories of the integer quantum Hall effect.

In order to evaluate the Hall conductance $G_H$ of the system, it is necessary to express $dI$ in terms of *electron* electro-chemical potentials, since these are the measured electro-chemical potentials of the contacts. This can be done using equation (4), which yields

$$d\left(\mu_1^* - \mu_2^*\right) = d(\mu_1 - \mu_2) - (m_b h/e)\, d\int_2^1 \mathbf{J}.\hat{\mathbf{B}} \times d\mathbf{r} = d(\mu_1 - \mu_2) + (m_b h/e)\, dI \tag{14}$$

Notice that only $m_b$, the bulk value of $m$, enters (14) since the contribution of the Type I modes to $dI$ is zero, as has been shown above. Combining (13) and (14), we find that for edge mode configurations of the type shown in Fig.2(b), the Hall conductance of the Hall bar in Fig.1 is

$$G_H = -e\frac{\partial I}{\partial(\mu_1 - \mu_2)} = \frac{e^2}{h}\frac{p}{m_b p - 1} \tag{15}$$

**Type III edge modes** are those that correspond to mean field composite fermion Landau levels that begin below the Fermi level deep in the interior of the two-dimensional conductor and pass upwards through the Fermi level as the edge of the sample is approached. This behavior is qualitatively similar to that of the usually discussed edge modes of the integer quantum Hall effect, which are in fact special cases of Type III edge modes. The two interior modes in Fig.2(a) are examples.

For Type III modes, equation (9) takes the form

$$dI_{m,r} = -\frac{e}{h}d\left(\mu_1^* - \mu_2^*\right). \tag{16}$$

and $dI$ is obtained by summing over the populated Type III modes to yield

$$dI = \sum_r dI_{m,r} = -p\frac{e}{h}d\left(\mu_1^* - \mu_2^*\right) \tag{17}$$

where $p$ is once again the number of occupied composite fermion Landau levels far from the edges of the Hall bar. Equation (14) applies to this case as well and, combining it with (17), we find that for edge mode configurations of the type shown in Fig.2(a), the Hall conductance of the Hall bar in Fig.1 is

$$G_H = -e\frac{\partial I}{\partial(\mu_1 - \mu_2)} = \frac{e^2}{h}\frac{p}{m_b p + 1}. \tag{18}$$

The factors $p/(m_b p \pm 1)$ that appear in the Hall conductance expressions (15) and (18) are equal to the values of the bulk electron Landau level filling parameter $\nu$ for the systems in which $p$ composite fermion landau levels are occupied in the bulk.[11,14] Thus the expressions (15) and (18) give the correct fractional Hall conductances $G_H = \nu e^2/h$ that are observed experimentally, and are in agreement with the results of Jain[11] for the corresponding Corbino systems. However, the above theory has been developed specifically to apply to experimental Hall bars with current and voltage contacts, and this is exploited in the following Sections, where the application to a number of different experiments is discussed.

The present theory can be thought of as the generalization to composite fermion systems of the edge channel theories of the integer quantum Hall regime.[4–8] However, there are fundamental qualitative differences between the present results in the fractional quantum Hall regime and those theories: As has been shown above, the present composite fermion theory implies the existence of "silent" edge channels that cross the Fermi level, but do not respond to a shift of the *quasi-equilibrium* electrochemical potential with a change in the electric current that they carry. Furthermore there are composite fermion modes separated from the Fermi level by a finite gap, that *do* respond with a change in the current that they carry to a change in the electrochemical potential. Another fundamental difference is that the effective edge electro-chemical potentials of the composite fermions differ from the electro-chemical potentials of the electron reservoirs, and *it is this difference that results in the Hall conductance being quantized in fractional (rather than integer) multiples of $e^2/h$.*

## 5. Transport Across Smooth Barriers and Through Adiabatic Constrictions

The edge state picture of the integer quantum Hall effect was greatly clarified by experiments in which different edge states were selectively back-scattered by obstacles in a Hall bar.[39] Analogous experiments have been performed in the fractional quantum Hall regime,[40,41,42] and have been interpreted in terms of phenomenological models of fractional edge channels,[43,44] that are not based on composite fermion theory. The experiments were carried out on Hall bars with potential barriers imposed across the current path by means of gates. Here we will consider the simplest such device, depicted in Fig.3: A 2DEG, partly covered by a gate, connecting two electron reservoirs at electro-chemical potentials $\mu_1$ and $\mu_2$. We will then apply the ideas that are developed to transport through adiabatic constrictions in the fractional quantum Hall regime, another class of devices that has been studied experimentally.[45]

In the case shown in Fig.3a, the 2DEG is in the integer quantum Hall regime at a Landau level filling factor $\nu = 1$, except under the gate where the electron density is depleted to $\nu = 2/3$. The arrows denote the directions of electron flow. The thickest black directed curves show where the $\nu = 1$ Landau level passes through the Fermi energy; these edge modes are reflected by the barrier that is set up by the gate. The darker shading denotes the regions occupied by Type I modes, most of which are not shown individually. The thinnest black directed curves (confined entirely to the shaded regions) show where $B_{eff} = 0$. The lighter shading indicates regions of very low density where a description as a pinned Wigner solid may be appropriate. The composite fermion Landau level structure along the line AC under the gate is that depicted in Fig.2b, for $m_b = 2$. The four black directed curves of intermediate thickness in Fig.3a show where the two Type II composite fermion Landau levels of the $\nu=2/3$ region rise up through the Fermi level as one leaves the region under the gate and the local electron density increases. Further away from the gate, these composite fermion Landau levels adiabatically become Type I edge modes of the $\nu = 1$ region.

We now consider the two-terminal conductance $G$ of the system in Fig.3a, assuming that the propagation of the composite fermion edge modes is adiabatic and dissipationless. Adiabatic propagation of fractional edge modes over the length scales of interest is supported by experimental data,[42,46] and should generally occur if the potential is slowly varying, as is assumed in this paper. Dissipationless transport is supported experimentally by the observation[40,47] of a longitudinal conductance plateau as a function of gate voltage at $G = 2/3\ e^2/h$, and our analysis applies in this regime.

Dissipationless transport implies that the effective composite fermion electro-chemical potential throughout the region in which the propagating edge modes originate *only* from the electron reservoir that is at $\mu_1$ ($\mu_2$) in Fig.3a is independent of position and equal to $\mu_1^*$ ($\mu_2^*$). In particular, the effective composite fermion electrochemical potential equals $\mu_1^*$ along the lines AB and BX and equals $\mu_2^*$ near C. Therefore we can apply the results obtained for Type II edge modes in Section 4 directly to obtain the net current $I$ crossing the line AC in Fig.3a. In this situation, equation (15) reduces to the result $G = -eI/(\mu_1-\mu_2) = 2/3\ e^2/h$ for the two terminal conductance of the device since $m_b = 2$ and $p = 2$ in the region under the gate. This result is in agreement with experiments.[40,47]

It may, at first sight, seem surprising that, in the present theory, a non-zero *net* current can flow through the barrier imposed by the gate in Fig.3a, since as has been mentioned above, the Type II modes that carry the current through the $\nu = 2/3$ region are the adiabatic continuation of Type I edge modes of the $\nu=1$ regions. (Recall that it was shown in Section 4 that Type I edge modes are "silent" modes and do *not* respond to a change in the quasi-equilibrium edge elec-tro-chemical potential with a change in the net current that they carry.)

This matter is clarified in Fig.4a, where we show the composite fermion energy level structure along the path A-B-X in Fig.3a. The two composite fermion Landau levels of interest are labelled s and s'. Notice that they begin at the apex of a fan in a region (on the line A-B) where the composite fermion effective electro-chemical potential equals $\mu_1^*$ and cross the Fermi level (on the line B-X) also at $\mu_1^*$. Therefore the argument given for Type I modes in Section 4 can be applied directly to these modes. The result is that the net currents that these modes carry across the line A-B-X in Fig.3a are independent of $\mu_1^*$, just like the currents carried by the Type I edge modes in the of the $\nu = 1$ region that feed the modes s and s'. The net current carried by s and s' across the line A-B-X can be viewed as consisting of two parts: The current that crosses AB (and carries electrons on to contact D in Fig.3a), and the current that crosses BX (and returns electrons to contact S). Thus when $\mu_1$ increases, the electron flux that s and s' carry across AB increases while that across BX decreases by the same amount. That is, although the electron flux carried by a Type I edge mode of the $\nu = 1$ region is independent of $\mu_1^*$, the way that this flux is *divided* at the gate into the components that continue to contact D and return to S depends on $\mu_1^*$. Because of this asymmetric branching at the potential barrier, some of the modes that are silent in the $\nu = 1$ region are able to transmit a current that depends on $\mu_1$ over the barrier, and therefore play a very important role in conduction through this system. Notice also that the electron flux after passing under the gate continues on to contact D, carried by Type I modes in the upper $\nu = 1$ region of Fig.3a. In this region the modes do *not* originate from a single contact and therefore are not described by a single well-defined quasi-equilibrium electro-chemical potential. Thus the Type I modes in this region are no longer silent. Such non-equilibrium effects will be discussed further in Section 6.

Another interesting case is shown in Fig.3b, where the 2DEG is at $\nu = 2/3$, except under the gate where $\nu = 1/3$. Here the Type II modes of the $\nu=2/3$ regions are reflected adiabatically at the barrier. The energy level structure along the line EG under the gate is similar to that shown in Fig.2a, for $m_b = 2$, but there is only one Type III composite fermion Landau level present ($p = 1$) instead of the two shown in Fig.2a. The thicker black curves in Fig.3b show where this Type III level passes through the Fermi energy under the gate; away from the gate this Landau level switches over adiabatically into Type I edge modes of the $\nu = 2/3$ regions. The areas occupied by Type I modes are again shown by the darker shading. Once again assuming adiabatic and non-dissipative transport, the net current $I$ across line EG can be obtained in a similar way to that across AC in Fig.3a, but applying the analysis given for Type III modes in Section 4. In this way, from equation (18) we obtain $G = -eI/(\mu_1-\mu_2) = e^2/(3h)$ for the two terminal conductance of the device, again in agreement with experiment.[40] The composite fermion Landau level structure along E-F-H is shown in Fig.4b. The important composite fermion Landau level in the $\nu=1/3$ region is labelled s. Notice that, as in Fig.4a, s begins at the apex of a fan and passes through the Fermi level at the same composite fermion effective electro-chemical potential $\mu_1^*$. Therefore the net current that it carries across the line E-F-H is independent of $\mu_1^*$, and the electron flux that is fed to it from the contact S by the silent mode (in the $\nu = 2/3$ region) branches asymmetrically in much the same way as the currents discussed above for the system in Fig.3a do.

It is straight forward to extend the above considerations to the case of any potential barrier with a Landau level filling parameter of the Jain form $\nu = p/(mp\pm1)$, across a 2DEG with any $\nu' = p'/(m'p'\pm1)$, that is, to other cases where both the 2DEG and the barrier are at densities for which the Hall conductance is quantized. (A suggestion for relevant experiments is made at the end of the Appendix.) These arguments are also readily extended to experimental geometries[40] in which a single gate covers a region of a Hall bar connected to several contacts and agreement with experi-ment is again obtained.

It should also be noted that the above treatment of 2D conductors with barriers also applies to *adiabatic* constrictions in which the potential is slowly varying with position. Such constrictions have been fabricated and studied experimentally in the fractional quantum Hall regime. For example, experimental studies have recently been reported[45] of a pair of parallel constrictions with $\nu = 1/3$ at their centers, in a $\nu = 2/3$ conductor, which exhibited an approximate conductance plateau near $G = e^2/(3h)$, modulated by weak Aharonov-Bohm oscillations that signal a small deviation from adiabaticity. The theory given above for the system in Fig.3b applies equally to individual constrictions of this type that are perfectly adiabatic. Constrictions that are strongly non-adiabatic have also been studied experimentally[48] in the fractional quantum Hall regime. The conductance of these structures was found to be dominated by strong resonances, and decreased as the temperature decreased, consistent with the predictions of chiral Luttinger liquid theory.[49] A discussion of the physics of such strongly non-adiabatic devices is beyond the scope of the present paper.

## 6. Selective Population and Detection of Composite Fermion Edge Channels: Non-Equilibrium Phenomena

It has been stressed in the literature[50] that an important test of theories of the fractional quantum Hall effect is provided by experiments[42] in which edge channels are selectively populated and detected. The experimental geometry is shown schematically in Fig.5 where the same notation is used as in Fig.3a. The current flows between contacts 1 and 4, the electron drain D and source S, respectively. The voltage contacts are 2 and 3; each of these draws no net current. This Hall bar is at $\nu = 1$, except under the two gates where $\nu = 2/3$. An important feature of this experiment was that the device was in the adiabatic regime: There was no appreciable inter-channel scattering taking place,[42] even along the 2 micron-long current path between the gate near contact 4 and that near contact 3, where edge states originating at contact 4 propagate side by side with those originating from contact 2 which is at a different electro-chemical potential. We analyze this system in the composite fermion edge channel picture as follows:

The situation along line C-E in Fig.5 is exactly like that along line A-C in Fig.3a, and the same analysis yields that the net current flowing out of contact 4 is $I = -2e(\mu_4 - \mu_2)/(3h)$. Consider now the line connecting contacts 4 and 3 in Fig.5, along which $B_{eff} = 0$ for $m=2$. This line is labelled "N." If transport is adiabatic (there is no inter-channel scattering), then along the segment of N from a to a' the carriers at the high energy edges of the two Type I modes that originate at contact 2 are out of equilibrium with those near the $B_{eff} = 0$ line N. The carriers near the $B_{eff} = 0$ line originate at contact 4 (most of these carriers belong to Type I modes that originate *only* at contact 4) and are thus at the effective electro-chemical potential $\mu_4^*$. That is, the effective electrochemical potential along the entire line N is $\mu_4^*$.[51] Given this, and the fact that the net current crossing line A-B is zero (since contact 3 is a voltage probe), it follows immediately that $\mu_3^* = \mu_4^*$, and hence $\mu_3 = \mu_4$. Since contact 2 is also a voltage probe, the net current crossing F-G is also zero, and therefore $\mu_2 = \mu_1$. It follows that the Hall conductance of the device in this configuration is $G_H \equiv -eI/(\mu_3 - \mu_2) = -eI/(\mu_4 - \mu_2) = 2/3 \ e^2/h$. This result is in agreement with the experimental data.[42]

Now consider the opposite limiting case where there is strong inter-channel scattering along the path from a to a' so that all of the modes traveling along the edge of the sample between the two gates equilibrate with each other completely. In that case the net current flowing out of contact 4 is still $I = -2e(\mu_4 - \mu_2)/(3h)$, as above. However, because of the equilibration, we can replace the sub-system consisting of contact 4 and the gate next to it with an equivalent ideal contact $x$ at an electro-chemical potential $\mu_x$, with $\mu_x$ chosen so that the net current flowing out of contact $x$ equals that flowing out of contact 4. This is achieved by setting $\mu_x - \mu_2 = 2(\mu_4 - \mu_2)/3$. But since contacts 2 and 3 are voltage contacts and no net current flows out of them, we have that $\mu_3 = \mu_x$, and $\mu_2 = \mu_1$. It then follows that $I = -2e(\mu_4 - \mu_2)/(3h) = -e(\mu_x - \mu_2)/h = -e(\mu_3 - \mu_2)/h$ and that $G_H \equiv -eI/(\mu_3 - \mu_2) = e^2/h$. The transition from the case discussed in the preceding paragraph where there is no mixing between edge modes (and $G_H = 2/3\ e^2/h$) to the present case of complete edge mode equilibration where $G_H = e^2/h$ has been observed experimentally.[52] Notice that the observation of such a transition is contrary to the predictions of the Luttinger liquid theories of edge states that have been proposed to date because in those theories a $\nu = 1$ edge supports only a *single* edge mode, so that a transition that requires the presence of more than one mode at the $\nu = 1$ edge cannot occur in those theories.

Now suppose that the current and voltage leads are interchanged, so that the current flows between contacts 2 and 3, and contacts 1 and 4 are the voltage probes and draw no current. The net current flowing out of contact 2 is then $I = -e(\mu_2 - \mu_1)/h$. Since contact 4 is a voltage probe, there is no net current across line C-E, and hence $\mu_2 = \mu_4$. The Hall conductance under these conditions is therefore given by $G_H' \equiv -eI/(\mu_4 - \mu_1) = e^2/h$ (irrespective of whether equilibration occurs along the line N in Fig.5), again in agreement with experiment.[42,52]

Thus the present composite fermion theory is able to account for the results of these experiments in which non-equilibrium fractional edge states play a key role. To our knowledge the only other theory that has been shown to yield results consistent with this experimental data is a phenomenology,[44] which is unrelated to composite fermion theory.

## 7. Conclusions

In this paper we have developed a mean field theory of composite fermion edge channel transport that yields the experimentally observed quantized Hall conductances in both the integer and fractional quantum Hall regimes. Three different types of composite fermion edge channels have been identified, including two types that are *qualitatively* different in character from the magnetic edge states discussed in previous theories. An *essential* role in the present theory is played by the effective composite fermion edge electro-chemical potentials which differ from those of ordinary electrons. This difference results in the fractional (as opposed to integer) quantization of the Hall conductance in the fractional regime. The behavior of the composite fermion edge states at adiabatic potential barriers and constrictions has also been analyzed, and shown to explain the results of experiments, including those in which these edge channels are selectively populated and detected and local quasi-equilibrium is violated.

Acknowledgment — This work was supported by the Natural Sciences and Engineering Research Council of Canada.

**Appendix: Comparison with Hartree Theories; Direction of Propagation of Composite Fermion Edge States and the Divergence of the Effective Mass.**

Recently Brey[29] and Chklovskii[30] have carried out numerical model calculations of composite Fermion edge state structure and energetics in the Hartree approximation for edges of finite width. Where $B_{\text{eff}} \approx 0$, they found behavior of the composite fermion Landau level energies $\varepsilon_{m,r}$ that was qualitatively different from that given by equation (5) and sketched in Fig.2. The difference is depicted schematically in Fig.6(a), which shows only the two composite fermion Landau levels that are populated in the interior of a $\nu = 2/3$ system. The curves labelled "II" indicate the behavior of the Type II modes discussed in the present work. They terminate at $\varepsilon = W$ where $B_{\text{eff}} = 0$. The corresponding composite fermion Landau levels of References 29 and 30 are shown by the thinner lines that, instead of terminating at $\varepsilon = W$, turn upwards and pass through the Fermi level.

As has been pointed out by Chklovskii,[30] this behavior of the levels found in the Hartree approximation can be understood in terms of the model of charged particles in a non-uniform magnetic field discussed by Müller.[53] For the situation in Fig.6(a), Müller's argument can be summarized briefly as follows. Inserting equations (6) and (8) in (7) yields

$$\left(-\frac{\hbar^2}{2m^*}\frac{\partial^2}{\partial x^2} + V_{\text{eff}}(x)\right)X_{k,m,r}(x) = \varepsilon_{m,r}(k)X_{k,m,r}(x) \tag{19}$$

where the effective potential is given by

$$V_{\text{eff}}(x) = \frac{(\hbar k + ebx^2)^2}{2m^*}. \tag{20}$$

Here for simplicity (following Müller) we set $B_{\text{eff}} = 2bx$ with $b > 0$ and ignore $W(x)$. In Fig.6(a), we are concerned with the case $x \leq 0$. Thus for $k < 0$, the minimum of $V_{\text{eff}}$ occurs at $x = -\sqrt{(-\hbar k/(eb))}$ and $\partial \varepsilon / \partial k < 0$ (the composite fermion velocity in the $y$-direction is negative). However, for $k > 0$ the minimum of $V_{\text{eff}}$ occurs at $x = 0$ and rises in energy as $k$ increases, so that $\partial \varepsilon / \partial k > 0$. That is, when the region where $B_{\text{eff}} \approx 0$ is approached from the left in Fig.6, the composite Fermion velocity in the y-direction changes from negative to positive and the energy begins to rise and passes through the Fermi level, as in References 29 and 30. This, however, disagrees with experiments[34] which show that the carriers at a $\nu = 2/3$ edge propagate in the same direction as non-interacting electrons, which in the present case is the *negative y*-direction.

On the other hand, the direction of propagation for the Type II modes of the present theory is the correct one, since for them $\varepsilon$ does not turn upwards (see Fig.6(a)) and $\partial \varepsilon / \partial k$ remains negative. This applies also to the Type I modes of the present theory that are shown in the region $x < 0$ in Fig.6(b) where $B_{\text{eff}} < 0$, and shows that $\partial \varepsilon / \partial k$ is negative for these modes as well. For the Type I modes shown in the region $x > 0$ in Fig.6(b) $\partial \varepsilon / \partial x > 0$, but $\partial x / \partial k < 0$ because $B_{\text{eff}} > 0$ for $x > 0$. (That $\partial x / \partial k < 0$ for these modes can also be seen by noting that these modes correspond to the minimum of $V_{\text{eff}}$ at $x = +\sqrt{(-\hbar k/(eb))}$ for $k < 0$.) Thus $\partial \varepsilon / \partial k$ is negative for these Type I modes as well. It is straight forward to extend the above argument for the Type I modes in the region $x > 0$ to the case of Type III modes at the same edge of the sample, since $B_{\text{eff}} > 0$ for Type III modes as well. It follows that these Type III modes also propagate in the negative $y$-direction. That is, at any particular edge of the sample, *all* of the modes of the present theory propagate in the same direction, the direction that is observed experimentally.

We suggest that the above disagreement between the results of References 29 and 30 and experiment may be due to the fact that those theories, being at the Hartree level of approximation, do not adequately reflect the many-body character of the states at $B_{\text{eff}} = 0$, which is manifested by the divergence of the composite fermion effective mass $m^*$ in uniform systems. A simple-minded way to think of this is as follows: References 29 and 30 take $m^*$ to be a constant, independent of $B_{\text{eff}}$. But if $m^*$ diverges strongly enough where $B_{\text{eff}} = 0$, both the effective potential energy term $V_{\text{eff}}$ given by equation (20) and the kinetic energy on the LHS of (19) should vanish where $B_{\text{eff}} = 0$. Thus the upturn of $\varepsilon$ near $B_{\text{eff}} = 0$ that is predicted by the argument that is presented immediately below equation (20) (and occurs in the Hartree approximation) should not happen. A more detailed analysis along these lines shows that if $m^*$ diverges more strongly than $(B_{\text{eff}})^{-2}$, then the local composite fermion Landau level energies obtained by solving equation (7) are equal to $W$ for states centered where $B_{\text{eff}} = 0$, as is assumed in the present paper. Recent experiments[17,18] suggest that the divergence of the effective mass may be as strong as $(B_{\text{eff}})^{-4}$ in *uniform* systems, while theories predict weaker singular behavior.[14] Thus while taking account of the divergence of the effective mass may make it possible to resolve this qualitative difficulty encountered by References 29 and 30 and to reconcile this aspect of the Hartree theories with the present work, further study of this problem is clearly required.

Another deficiency of the Hartree theories of composite fermion edge structure[29,30] is that, unlike the present theory, they can not treat edges at which composite fermion Landau levels that derive from different values of $m$ are present simultaneously. This is a serious shortcoming of the Hartree theories at all values of the bulk filling fraction $\nu$, but it is especially severe when $\nu = 1$ or $1/3$ in the bulk. For instance, in order to model a $\nu = 1$ edge, it is assumed in Ref. 30 that *all* of the composite fermions (including those in the bulk) are of the $m=2$ type. This makes it possible to model a $\nu = 1$ edge at which the effective magnetic field passes through zero when the electron density passes through $\nu = 1/2$, but at the expense of the composite fermion effective magnetic field in the $\nu = 1$ bulk pointing in the direction *opposite* to that of the true magnetic field. This is problematic since it would imply that the energy gap between bulk composite fermion Landau levels decreases as the true magnetic field increases, whereas the opposite behavior is normally observed experimentally in the integer quantum Hall regime; the $m = 0$ composite fermion picture is the one that correctly describes the $\nu = 1$ bulk under most conditions. Similarly, for the $\nu = 1/3$ edge the Hartree models have to choose between the $m = 2$ and $m = 4$ pictures, neither of which describes *both* the bulk and the edge composite fermion energy level structure satisfactorily at the Hartree level of approximation. By contrast, we assume that the *local* composite fermion energy level structure is everywhere that associated with the *local* value of $m$ that conforms in turn to the local value of $\nu$, so that the above difficulties do not arise in the present work.

Finally, we note that in the present work the edge potential $W(x)$ is assumed to be very slowly varying. The numerical results presented in References 29 and 30 suggest to us that the number of silent Type I modes present at an edge may decrease as the slope of the edge potential increases. It would be of considerable interest to investigate this experimentally by fabricating samples with controllable potential profiles at the edges, and crossed by gates as in Section 5: If a larger number of silent edge modes is present in the system for smoother edge potentials, then a larger number of distinct fractional plateaus should be observable experimentally in the two-terminal conductance as the barrier voltage is varied.

**References:**

[1] K. von Klitzing, G. Dorda and M. Pepper, Phys. Rev. Lett. **45**,494 (1980).

[2] D. C. Tsui, H. L. Störmer and A. C. Gossard, Phys. Rev. Lett. **48**,1559 (1982).

[3] R. B. Laughlin, Phys. Rev. B**23**, 5632 (1981).

[4] P. Streda, J. Kucera and A. H. MacDonald, Phys. Rev. Lett. **59**,1973(1987).

[5] J. K. Jain and S. A. Kivelson, Phys. Rev. Lett. **60**,1542 (1988).

[6] M. Büttiker, Phys. Rev. B**38**, 9375 (1988).

[7] R. Landauer, IBM J. Res. Dev. **1**, 223 (1957); M. Büttiker, Phys. Rev. Lett. **57**, 1761 (1986).

[8] B. I. Halperin, Phys. Rev. B**25**, 2185 (1982).

[9] R. B. Laughlin, Phys. Rev. Lett. **50**, 1395 (1983).

[10] F. D. M. Haldane, Phys. Rev. Lett. **51**, 605 (1983); B. I. Halperin, Phys. Rev. Lett. **52**, 1583 (1984); R. B. Laughlin, Surf. Sci. **141**, 11 (1984).

[11] J. K. Jain, Phys. Rev. Lett. **63**, 199 (1989); Phys. Rev. B**41**, 7653 (1990); Science **266**, 1199 (1994).

[12] F. Wilczek, Phys. Rev. Lett **49**, 957 (1982); F. Wilczek and A. Zee, Phys. Rev. Lett **51**, 2250 (1983); D. Arovas, J. R. Schrieffer and F. Wilczek, Phys. Rev. Lett **53**, 772 (1982); S. M. Girvin and A. H. MacDonald, Phys. Rev. Lett. **58**, 1252 (1987); R. B. Laughlin, Phys. Rev. Lett. **60**, 2677 (1988).

[13] A. Lopez and E. Fradkin, Phys. Rev. B**44**, 5246 (1991).

[14] B. I. Halperin, P. A. Lee and N. Read, Phys. Rev. B**47**, 7312 (1993).

[15] R. L. Willet, J. P. Eisenstein, H. L. Störmer, D. C. Tsui, A. C. Gossard and J. H. English, Phys. Rev. Lett **59**, 1776 (1987).

[16] R. R. Du, H. L. Störmer, D. C. Tsui, L. N. Pfeiffer and K. W. West, Phys. Rev. Lett **70**, 2944 (1993).

[17] R. R. Du, H. L. Störmer, D. C. Tsui, A. S. Yeh, L. N. Pfeiffer and K. W. West, Phys. Rev. Lett **73**, 3274 (1994); I. V. Kukushkin, R. J. Haug, K. von Klitzing, and K. Ploog, Phys. Rev. Lett **72**, 736 (1994).

[18] H. C. Manoharan, M. Shayegan and S. J. Klepper, Phys. Rev. Lett **73**, 3270 (1994).

[19] R. L. Willet, R. R. Ruel, K. W. West and L. N. Pfeiffer, Phys. Rev. Lett **71**, 3846 (1993).

[20] W. Kang, H. L. Störmer, L. N. Pfeiffer, K. W. Baldwin and K. W. West, Phys. Rev. Lett **71**, 3850 (1993).

[21] V. J. Goldman, B. Su and J. K. Jain, Phys. Rev. Lett **72**, 2065 (1994).

[22] B. Rejaei and C. W. J. Beenakker, Phys. Rev. B**46**, 15566 (1992).

[23] B. I. Halperin, Lecture presented at the 22nd International Conference on the Physics of Semiconductors, Vancouver, Canada, August, 1994.

[24] A. S. Goldhaber and J. K. Jain, 1995, paper cond-mat/9501080 at the e-print archive cond-mat@xxx.lanl.gov or the www site http://xxx.lanl.gov/, Physics Letters A**199**, 267 (1995).

[25] S. C. Zhang, H. Hansson and S. A. Kivelson, Phys. Rev. Lett. 62, 82 (1989); S. C. Zhang, Int.

J. Mod. Phys. B**6**, 25 (1992).

[26] B. Rejaei and C. W. J. Beenakker, Phys. Rev. B**43**, 11392 (1991).

[27] B. I. Halperin, Phys. Rev. B**45**, 5504 (1992).

[28] D. R. Leadley, R.J.Nicolas, C. T. Foxon and J. J. Harris, Phys. Rev. Lett **72**, 1906 (1994).

[29] L. Brey, Phys. Rev. B**50**, 11861 (1994)

[30] D. B. Chklovskii, Phys. Rev. B**51**, 9895(1995), paper cond-mat/9502014 at the e-print archive cond-mat@xxx.lanl.gov or the www site http://xxx.lanl.gov/

[31] In particular, this result applies to the composite fermion edge modes in the $\nu = 2/3$ regime, the case depicted in Fig.2(b) for $m_b = 2$. In this respect the present theory differs from the fractional edge channel theory of MacDonald[32] and from the chiral Luttinger liquid theory of Wen,[33] which are not based on the composite fermion picture. Both of these theories predict the occurrence of two edge modes, propagating in *opposite* directions, at the same $\nu = 2/3$ edge, behavior that has not been detected experimentally.[34] Meir has suggested recently [35] that the theory of MacDonald is applicable to steeply sloping confinement potentials, but that the nature of the edge channels changes qualitatively when the potential becomes smoother. This may be relevant, since the present theory assumes the confining potential to be *slowly* varying. Taking a different approach, Kane, Fisher and Polchinski[36] have recently shown within chiral Luttinger liquid theory that counter-propagating edge modes should not manifest themselves directly in the experiments of Ashoori *et al.*[34] if random defect scattering is present, behavior reminiscent of that predicted[37] for counter-propagating modes in weakly disordered non-interacting chiral electron systems.

[32] A. H. MacDonald, Phys.Rev. Lett. **64**, 220 (1990).

[33] X. G. Wen, Phys.Rev. Lett. **64**, 2206 (1990); Phys. Rev. B**43**, 11025 (1991); Phys. Rev. B**44**, 5708 (1991).

[34] R. C. Ashoori, H. L. Störmer, L. N. Pfeiffer, K. W. Baldwin, and K. West, Phys. Rev. B**45**, 3894 (1992).

[35] Y. Meir, Phys.Rev. Lett. **62**, 2624 (1994).

[36] C. L. Kane, M. P. A. Fisher and J. Polchinski, Phys. Rev. Lett. **72**, 4129 (1994).

[37] C. Barnes, B. L. Johnson and G. Kirczenow, Phys. Rev. Lett. **70**,1159 (1993); Canadian Journal of Physics **72**, 559 (1994).

[38] Note that the real magnetic field **B** is assumed to be spatially uniform.

[39] R. J. Haug, A. H. MacDonald, P. Streda and K. von Klitzing, Phys. Rev. Lett. **61**, 2797 (1988); S. Washburn, A.B.Fowler, H. Schmid and D. Kern, Phys. Rev. Lett. **62**, 1181 (1989); B. J. van Wees, E. M. M. Willems, C. J. P. Harmans, C. W. J. Beenakker, H. van Houten, J. G. Williamson, C. T. Foxon and J. J. Harris, Phys. Rev. Lett. **61**, 2801 (1988); S. Komiyama, H. Hirai, S. Sasa, and S. Hiyamizu, Phys. Rev. B**40**, 12566 (1989); R. J. Haug and K. von Klitzing, Europhys. Lett. 10, 489 (1989); B. W. Alphenaar, P.L.McEuen, R. G. Wheeler and R. N. Sacks, Phys. Rev. Lett. **64**, 677(1990).

[40] A. M. Chang and J. E. Cunningham, Solid State Commun. **72**, 651 (1989).

[41] R. J. Haug, K. von Klitzing, K. Ploog and P. Streda, Surface Science **229**, 229 (1990).

[42] L. P. Kouwenhoven, B. J. van Wees, N. C. van der Vaart, C. J. P. M. Harmans, C. E. Timmering and C. T. Foxon, Phys.Rev. Lett. **64**, 685 (1990).

[43] A. M. Chang, Solid State Commun. **74**, 871 (1990).

[44] C. W. J. Beenakker, Phys.Rev. Lett. **64**, 220 (1990).

[45] C. J. B. Ford, P. J. Simpson, I. Zailer, J. D. F. Franklin, C. W. H. Barnes, J. E. F. Frost, D. A. Ritchie and M. Pepper, Journal of Physics: Condensed Matter **6**, L725 (1994). See also V. J. Goldman and B. Su, Science **267**, 1010 (1995).

[46] J. K. Wang and V. J. Goldman, Phys.Rev. Lett. **67**, 749(1991); Modern Physics Letters **5**, 1617(1991).

[47] L. P. Kouwenhoven *et al.*, see Fig. 96 p. 210 in C. W. J. Beenakker and H. van Houten, Solid State Physics **44**,1(1991).

[48] F. P. Milliken, C. P. Umbach and R. A. Webb, unpublished; see Physics Today Vol. **47**, No 6, p. 21 (1994).

[49] K. Moon, H. Yi, C. L. Kane, S. M. Girvin and M. P. A. Fisher, Phys. Rev. Lett **71**, 4381 (1993).

[50] C.W. J. Beenakker and H. van Houten, Solid State Physics **44**,1(1991).

[51] It is worth noting here that since, on the line N, the direction of **J** is parallel to the line N, equation (3) implies that $\int \mathbf{E}_g \cdot d\mathbf{r}$ along line N is zero, so that the gauge electric field does not result in electro-chemical potential differences along line N.

[52] A. M. Chang and J. E. Cunningham, Phys. Rev. Lett **69**, 2114 (1992).

[53] J. E. Müller, Phys. Rev. Lett. **68**, 385 (1992).

**Figure Captions:**

**Fig.1** Schematic drawing of a two-dimensional conductor connected to electron source and drain contacts S and D at electrochemical potentials $\mu_1$ and $\mu_2$, respectively. The directions of electron flow at the edges of the sample are shown by arrows. L and R are Hall voltage contacts which draw no net current. The composite fermion Landau level structure along the dashed line is shown in Fig.2.

**Fig.2** Schematic representation of the mean field composite fermion energy levels $\varepsilon_{m,r}$ (the thinner lines) and of the self-consistent confining potential $W$ (the heavier lines) as a function of position along the dashed line in Fig.1. (a) and (b) are examples in which the effective magnetic field in the bulk is parallel and anti-parallel to the real magnetic field, respectively. Examples of Type I, II and III composite fermion edge modes are as indicated. $\mu_1^*$ and $\mu_2^*$ are the effective composite fermion electro-chemical potentials at edges 1 and 2 (as labelled in Fig.1), respectively.

**Fig.3** (a) A 2D conductor in the $\nu = 1$ integer quantum Hall regime traversed by a gate under which the 2DEG is depleted to the $\nu = 2/3$ fractional regime. (b) A 2D conductor in the $\nu = 2/3$ fractional quantum Hall regime traversed by a gate under which the 2DEG is depleted to the $\nu = 1/3$ fractional regime. Arrows indicate the direction of electron flow. Dark shading indicates regions occupied by Type I composite fermion modes. The thickest black directed curves show where the $\nu = 1$ Landau level passes through the Fermi energy. The thinnest black directed curves show where $B_{\text{eff}} = 0$. The black directed curves of intermediate thickness show where some of the $m = 2$ composite fermion Landau levels pass through the Fermi energy.

**Fig.4** Composite fermion Landau levels (thin lines) and potential energy function W (thick lines) plotted schematically as a function of position (a) along the path A-B-X in Fig.3a and (b) along the path E-F-H in Fig.3b.

**Fig.5** A 2D conductor in the $\nu = 1$ integer quantum Hall regime traversed by two gates under which the 2DEG is depleted to the $\nu = 2/3$ fractional regime. Notation as in Fig.3. When a current flows between contacts 1 and 4, and contacts 3 and 2 are voltage probes that draw no current, the adiabatic edge channels near the path a - a' are not in quasi-equilibrium with each other. This results in an anomalous observed[42] Hall conductance $2e^2/(3h)$, which is explained by the present composite fermion theory; see text.

**Fig.6** (a): Schematic drawing of the behavior of the two composite fermion Landau levels that are populated in the interior of a $\nu = 2/3$ system. The curves labelled "II" indicate the behavior of the Type II modes discussed in the present work. The behavior of the corresponding composite fermion Landau levels found in References 29 and 30 is shown as the thinner lines that, instead of terminating at $\varepsilon = W$, turn upwards and pass through the Fermi level. (b): As in Fig.6(a) but showing the Type I modes of the present theory that correspond to $m=2$ and not the modes of References 29 and 30. The location where $B_{\text{eff}} = 0$ for $m = 2$ and $x = 0$ (in the notation of the Appendix) is also indicated.